\documentclass[hyper,a4paper]{JHEP3}
\usepackage{multirow}

\usepackage[dvips]{graphicx}
\usepackage{epsfig}
\usepackage{subfigure}
\usepackage{rotating}
\usepackage{amssymb}
\usepackage{amsmath}
\def\lesssim{\mathrel{\hbox{\rlap{\hbox{\lower5pt\hbox{$\sim$}}}\hbox{$<$}}}}
\def\gtrsim{\mathrel{\hbox{\rlap{\hbox{\lower5pt\hbox{$\sim$}}}\hbox{$>$}}}}

\def\bea{\begin{eqnarray}}
\def\eea{\end{eqnarray}}

\def\mgluino{m_{\tilde{g}}}

\def\ptmiss{\not\!\!{p_T}}

\def\beq{\begin{equation}}   %
\def\eeq{\end{equation}}   %

\title{Production of two Higgses at the Large Hadron Collider 
in CP-violating MSSM}

\author{Priyotosh Bandyopadhyay$^{a,1}$ and Katri Huitu$^{a,2}$\\

$^a$Department of Physics,  and Helsinki Institute of Physics,\\
P.O.Box 64 (Gustaf H\"allstr\"omin katu 2), FIN-00014 University of Helsinki, Finland\\
Email: \email{$^1$priyotosh.bandyopadhyay@helsinki.fi, $^2$katri.huitu@helsinki.fi}}

\abstract{Production of two Higgs bosons is studied in a
CP violating supersymmetric scenario at the 
Large Hadron Collider with $E_{cm}=14$ TeV. There exists a region
where a very light Higgs $\lesssim 50$ GeV could not be probed by
 LEP experiment. This leads to so called 'LEP hole' region. 
Recently LHC found a Higgs boson around $\sim 125$ GeV,
which severely constrains the possibility of having lighter Higgs bosons,
which cannot be detected, i.e., buried Higgs, in this model. We investigate
 the possibility of buried Higgs bosons 
along with the direct and indirect bounds coming from LEP, LHC and other
 experiments. In particular we take into account the constraints from EDM and
from $B$-observables. We analyse first the case where
a Higgs boson mass is around 125 GeV and the other two Higgs masses  are 
below 100 GeV and not observabed so far. In the second case the lightest Higgs boson 
mass is around 125 GeV and the other two are decoupled. We analyse the
production of two Higgses and their decay modes leading to various final
 states for these benchmark points. We perform a collider simulation 
with PYTHIA and Fastjet where we consider all the major backgrounds.
Among the final states we have analysed, we find that $2b+2\tau$ is
promising and the signal significance is  $5\sigma$ at an integrated 
luminosity $\lesssim 10$ fb$^{-1}$. For some benchmark points it is also
 possible to observe the light Higgs mass peak. We also explore the leptonic final state which could be instrumental in the precision measurement of a very light Higgs. }

\keywords{Higgs, CP-violation, Supersymmetry, LHC}
\preprint{KIAS P11035\\
   HIP-2011-18/TH}

\begin{document}
\section{Introduction}
 
CP violation is among the phenomena which are not fully understood in the context of the Standard Model (SM).
Although CP violation exists in the SM, and agrees well with the laboratory experiments, there is an inconsistency 
between the amount of violation and matter content of the Universe, and it is argued that new sources 
of CP violation are needed.

Many of the proposals for beyond the SM (BSM) physics do contain new sources for CP violation. In this work we consider Minimal Supersymmetric Standard Model (MSSM).
 It has been shown in the literature that the tree-level
CP invariance of the MSSM Higgs potential can be violated by loop effects involving CP-violating interactions of Higgs bosons to top and bottom squarks \cite{cpv,cpv1,cpv2,cpv3,cpv4,cpv5}. In such a scenario with explicit CP-violation at tree-level, the neutral Higgses ($h_i$, i=1,2,3) mix the CP states at loop-level.
It has been shown that \cite{cpv1,cpv3,cpv4} loop-induced CP-violation 
modifies the tree-level Higgs coupling such that light Higgs in this scenario could be $\lesssim 60$ GeV and can escape the detection at LEP2. 
 
For example, it has been shown that assuming universality of gaugino masses ($M_i$, i=1,2,3) at some high scale and assuming corrections from third generation strong sector, the  CP-violating MSSM Higgs sector can be parametrised in terms of a few independent phases \cite{cpphase}: the phase of Higgsino mass parameter
 (also called $\mu$ term), i.e., Arg($\mu$), and the phase of soft trilinear 
supersymmetry (SUSY) breaking parameters, i.e., Arg($A_f$), with $f=t, b$.
The experimental upper bounds on the electric dipole moments (EDMs) of
 electrons and  neutrons \cite {edm1, edm2} as well as of mercury
 atoms \cite{edm3} constrain these phases.

Before Large Hadron Collider (LHC) found out a Higgs resonance with mass around
125 GeV \cite{Higgsd1,Higgsd2,cms2g,atlas2g} earlier colliders had given bounds
on the Higgs mass. For the SM Higgs boson, mass bound from the LEP collider is $m_h>114.4$ GeV \cite{Barate:2003sz,Schael:2006cr}, and Tevatron excludes Higgs 
for the mass ranges $m_h\sim 147-180$ GeV and $100-103$ GeV but finds an excess in
115-135 GeV region \cite{tevatron}. In the MSSM with real and CP-conserving parameters, the experimental lower limit on the lightest Higgs boson is $\sim$ 90 GeV \cite{susylim} for any $\tan{\beta}$. 
The lower bound on the mass of the lightest Higgs boson of  
the CP-conserving MSSM  from LEP \cite{Schael:2006cr} can be drastically reduced 
or may even entirely vanish if non-zero CP-violating phases are allowed 
\cite{Bechtle:2006iw,Carena:2000ks}. This can happen through radiative corrections to 
the Higgs potential, whereby the above mentioned phases of the $\mu$ parameter 
and the $A$ parameters enter into the picture \cite{cpv,tdpl}.

 With the discovery of $\sim 125$ GeV Higgs at the LHC the question of a buried Higgs remains to be answered. The LHC experiment will look in all possible different decay 
modes to find an extra scalar which would be lighter than 100 GeV. Finding of
 such scalar(s) will be certainly a proof of BSM Higgs but also the possibility of CP-violating MSSM  will come into the picture. The phenomenology
of such a light Higgs has been studied in the context of CPV-MSSM in a benchmark scenario known as 'CPX'  \cite{cpv,tdpl}.

In the CPX scenario the $ZZh_1$ coupling can be strongly reduced because of
the CP violating phases, and the LEP mass limit for the lightest Higgs 
boson can be lowered to 50 GeV or even less, depending on $\tan\beta$.
Thus the LEP searches leave a hole in $(m_{h_1},\tan\beta)$ parameter space \cite{Schael:2006cr}. Complementary channels such as $e^+e^-\to h_1h_2$ suffer also phase space suppression within the hole region. At Tevatron, this CP violation and the Higgs phenomenology has been studied \cite{Carena:2002bb, Das:2010ds}.

 Within the hole region in addition to $ZZh_1$ coupling, $WWh_1$ and $tth_1$ are suppressed and thus the lightest Higgs boson $h_1$ is difficult to discover.
In the context of CPX scenarios there has been quite a few studies performed 
in the SM production channels \cite{Accomando:2006ga} as well as in the
supersymmetric channels \cite{Bandyopadhyay:2010tv,Li:2006hq,Bandyopadhyay:2007cp}. In the context of CP-conserving MSSM, cascade Higgs production has been
 studied  in \cite{Bandyopadhyay:2008sd}.

Most of these earlier studies do not fit with the data for $\sim 125$ GeV Higgs 
and the other experimental constraints coming from EDMs and the rare $B$-decays.
In this article we consider the recent SUSY mass bounds from LHC along with the Higgs results. We take into account thallium EDM result and constraints coming from Br($B_s\to s \gamma$) and Br($B_s\to \mu \mu$). We look for the possibility of the buried Higgs or the decoupled Higgs scenarios as two possibilities. In this context we study the Higgs pair production. We consider the $H\to b\bar{b}, \tau\bar{\tau}, \ell\bar{\ell}$ decay modes for possible final states. We find
that $2b+2\tau$ channel is very promising in searching for a very light Higgs ($m_{h_1}\sim 30$ GeV) and $\lesssim 10\, \rm{fb}^{-1}$ of integrated luminosity will
be enough to have $5\sigma$ significance over the dominant SM backgrounds. For
the precision measurement leptonic channel would be crucial. We also find out that Higgs productions in association with $Z$ will also contribute to these final states. One can differentiate between the two types of contributions by
constructing the heavier Higgs mass peak, i.e., $m_{h_{2,3}}$ in the 
corresponding channels at very high luminosity.

We will also point out that in certain benchmark points the two Higgs 
production through coupling of three Higgs bosons is important, and thus we have a possibility to probe the Higgs potential at those points. Obviously construction of the Higgs potential would be of fundamental importance.

The paper is organised as follows. In Section 2 we review the 
CPX benchmark scenario and discuss the experimental constrains.
 We also discuss very briefly the possibilities of evading such bounds.  
In section 3 we define the benchmark points consistent with the
experimental results for the collider study. The corresponding production 
cross-section and the decay branching fractions are listed in this section. 
In section 4 we carry out collider simulation for 14 TeV LHC for the desired 
final states. Finally in section 5 we summarise.

\section{CP violating scenario and the experimental constraints} 
It is known  \cite{cpv,tdpl} that the $CP$-mixing term in the Higgs sector
 is generated at quantum level and proportional to $Im(\mu A_t)/M^2_{SUSY}$. The well known CPX scenario predicts that  certain parameters are related:
\bea
&&m_{\tilde{t}} = m_{\tilde{b}} = m_{\tilde {\tau}} = M_{SUSY},\,\,
|A_t| = |A_b| = |A_{\tau}| = 2 M_{SUSY}, \nonumber\\
&&arg(A_{t})=arg(A_{b})=arg(A_{\tau})=90^0.
\eea

In particular the parameter space with $M_{SUSY}=500\,{\rm GeV}$ 
is of special phenomenological interest along with the other parameters that are compatible with the LEP ``hole'' and are  given below,

\bea
&&M_{SUSY}=500\,{\rm GeV},\,\,\quad |\mgluino|= 1\,{\rm TeV},\,\, 
M_2 = 2 M_1 = 200\,{\rm GeV},\,\,\quad\nonumber\\
&&arg(A_{b,\tau}) = 90^\circ,\,\,\quad arg(\mgluino) = 90^\circ ,\,\,
\tan\beta= 5 - 10.
\eea
In addition, $\tan\beta$ and $m_{H^\pm}$ are the free input parameters that could
be varied to achieve various points in the 'LEP hole'. The consequences of the CPX scenario have been studied in \cite{CPVstudies}.

Recently new results from LHC have changed the scenario as most of 
the parameter region is ruled out. In this paper we shall take into
account the Higgs discovery around $\sim 125$ GeV which has been reported by the CMS and ATLAS collaborations \cite{Higgsd1,Higgsd2}. Along with the recent LHC Higgs results we also consider the Higgs bounds from LEP \cite{leph}. We can see that buried Higgs, i.e., a very light Higgs ($\lesssim 60$ GeV), is still possible in MSSM. The possibility of a light Higgs could be an artifact of explicit CP-violation in the Lagrangian and then a loop-induced CP-violation in the Higgs sector as explained in the introduction. 

The recent studies on some indirect variables show that they can constrain
these CP-violating phases and eventually can rule out a large amount of parameter space. The bounds on the CP-violating MSSM coming from various dipole-moment measurements have been studied in details \cite{edmstudy}. In this paper we 
consider the constraints coming from electric-dipole moment (EDM) of thallium with the current $2\sigma$ upper bound  $|d_{Tl}|<1.3\times 10^{-24}$ e cm \cite{Thledm}. For this purpose we vary the relative angles 
between $M_1$, $M_2$ and also $\phi_{A_t}$, $\phi_{M_3}$; where we denote 
$Arg(A_f)=\phi_{f}$ and $Arg(M_i)=\phi_{M_i}$. In this region the one loop-SUSY contribution and light Higgs mediated two-loop contribution are comparable and 
tend to cancel each other. Thus it is possible to achieve the desired EDM bounds.
 Here we would like to mention that a very light Higgs ($m_{h_1}\lesssim 8$ GeV) is 
ruled out from bottomonium decay $\Upsilon(1S)\to \gamma h_1$ \cite{h1bnd}.

We also look into the flavour constraints coming from the $B$-observables. For this 
purpose we first consider Br($B_s\to \mu \mu$), which recently has come down by two
orders of magnitude \cite{bsmumu} can severely constrain this scenario. Br($B_s\to \mu \mu$) grows large as $\tan{\beta}$ increases. For the cancellation we use GIM operative point mechanism \cite{GIM}: we vary $\rho=\frac{Q_{1,2}}{Q_3}$, the ratio of first two generation of the squark masses over the third generation squark masses. The cancellation happens when $\rho \sim 0.8-1.9$. This predicts very light first two generation masses for some cases. To evade such light mass bound coming from $jets + \ptmiss$ at the LHC \cite{djlhc},  LSP mass must be 
large which would make the jets rather soft.

Next we consider the bounds coming from Br($B_s\to X_s \gamma$) \cite{bsg}. Unlike $B_s\to \mu \mu$ case Br($B_s\to X_s \gamma$) decreases as $\tan{\beta}$ increases. This is because the charged Higgs contribution is suppressed due to the threshold corrections at large $\tan{\beta}$ \cite{thcor}. We also included recent bounds on 
third-generation squark masses and on LSP from 8 TeV LHC \cite{thdgensusy}. To have light third-generation mass ($M_{\rm{SUSY}}$), the LSP needs to be relatively heavy, i.e., around $300$ GeV \cite{thdgensusy}.  We  also choose $m_3=1.4$ TeV to satisfy recent gluino mass bound \cite{djlhc,glnbd}. For this choice of gluino mass we find that it is very difficult to get $m_{h_3}\gtrsim 124$ GeV by using CPsuperH \cite{cpsuph2.0}\footnote{There is $\sim 2-3$ GeV uncertainty in Higgs mass calculated by CPsuperH and FeynHiggs \cite{FeynH}. For this paper we have used CPsuperH2.0 for the mass spectrum and the other observables.}. We vary $\tan\beta$ and $m_{H^\pm}$ as usual as we move to different points in the 'LEP hole'. The Higgs mass spectrum depends on 
the radiative correction which is very sensitive to top mass. The central value of $m_t$ has shifted frequently during the years. These shifts change the size of the hole,  although the location remains almost the same.  We use for the 
top mass $173.2 \pm 0.9$ GeV as referred by Tevatron \cite{topmass}.  

\section{Benchmark points for collider study}
After above investigation we find three points in explicit
CP-violating MSSM which are no longer so called ``CPX''
points but experimentally allowed ones. Allowed regions of 
the parameter space have very different but attractive phenomenological
consequences. Table~\ref{tabbps} describes the benchmark points that we consider 
 for our collider study. We consider three different scenarios:

\begin{enumerate}
\item Two light Higgses are buried and have masses $< 100$ GeV and mass of the
 heaviest one is around 125 GeV.
\item The lightest Higgs is very light $m_{h_1}\leq 30$ GeV, so that $h_3\to h_1 Z$ is allowed. The second lightest is also buried, $m_{h_2}\leq 100$ GeV and mass of the third one 125 GeV as in the previous case. 

\item This is a decoupled scenario where the heavier Higgses are decoupled with masses $\geq 500$ GeV and the lightest one has $m_{h_1}\sim 125$ GeV. 
\end{enumerate}

\begin{table}
\renewcommand\baselinestretch{0.2}
\begin{center}
\begin{tabular}{||c|c|c|c|c|c||}
\hline\hline
Parameters&BP1&BP2&BP3\\
\hline\hline
$\tan{\beta}$&30&30&20\\
\hline
$m_{H^\pm}$&115& 115&500\\
\hline
$\mu$ &1400&2000&1000\\
\hline
$M_1$ &300&300&300\\
$\phi_{M_1}$ &66&66&40\\
\hline
$M_2$ &400&400&400\\
$\phi_{M_2}$ &0&0&0\\
\hline
$M_3$ &1400&1400&1400\\
$\phi_{M_3}$ &61&61&60\\
\hline
$A_t$ &1000&1000&1000\\
$\phi_{A_t}$ &60&60&60\\
\hline
$A_b$ &11200&4200&1000\\
$\phi_{A_b}$ &35&35&90\\
\hline
$A_{\tau}$ &14200&16100&1000\\
$\phi_{A_{\tau}}$ &90&90&90\\
\hline
$\rho$ &0.83&0.88&1.90\\
\hline
\hline
$m_{h_1}$& 54.25&25.00 &123.50 \\
\hline
$m_{h_2}$& 95.00 & 94.70 &490.70 \\
\hline
$m_{h_3}$& 124.40 & 124.60&494.70 \\
\hline\hline
\end{tabular}
\vspace*{0.0mm}
\caption{Input parameters in the benchmark points within the 
'LEP-hole' and the corresponding CP-violating neutral Higgs masses. The angles are given in the unit of degree and other parameters are in GeV except $\tan{\beta}$ which is unitless.}\label{tabbps}
\end{center}
\end{table}

In this study, we focus on two Higgs production processes, i.e., $h_ih_j$, i=1,2,3 and j=2,3. We investigate the various possible decays of the Higgs bosons which will
lead to the corresponding final states. We also include Higgs production in association with a $Z$ boson. Table~\ref{crossH} presents the cross-sections of two Higgs boson productions ($h_ih_j, i,j=1,2,3$) for the three benchmark points at the LHC with ECM=14 TeV and Table~\ref{crossHZ} presents the cross-sections of Higgs boson productions associated with a $Z$ boson for the center of mass energy of 14 TeV.

\renewcommand{\arraystretch}{1.0}
\begin{table}
\begin{center}
\begin{tabular}{||c||c|c|c|c|c|c||}
 \hline

\hline
Benchmark &\multicolumn{6}{|c|}{Cross-section in fb}\\
Points &$\sigma_{h_{1} h_2}$ &$\sigma_{h_{1} h_3}$&$\sigma_{h_{1} h_1}$&$\sigma_{h_2h_2}$&$\sigma_{h_3h_3}$&$\sigma_{h_2h_3}$\\
\hline
\hline
BP1&908.02&47.02&5393.50&24.11 &7.83&6.92\\
\hline
BP2 &1858.89  &45.23&33086.7&20.35& 5.19&3.91 \\
\hline
BP3&$1.73\times 10^{-2}$& $1.0\times 10^{-2}$&18.6&$8.6\times 10^{-3}$& $5.7\times 10^{-3}$&0.47\\
\hline
 \hline
\end{tabular}
 \caption{Cross-sections (in fb) of two Higgs productions ($h_{2,3} h_i=1,2,3$)
at the LHC with $E_{cm}=14$ TeV for the benchmark points.}
\label{crossH}
\end{center}
\end{table}

\renewcommand{\arraystretch}{1.0}
\begin{table}
\begin{center}
\begin{tabular}{||c||c|c|c|c|c|c||}
 \hline

\hline
Benchmark &\multicolumn{3}{|c|}{Cross-section in fb}\\
Points &$\sigma_{h_{1}Z}$&$\sigma_{h_{2}Z}$&$\sigma_{h_{3} Z}$\\
\hline
\hline
BP1&513.18 &155.39&672.74\\
\hline
BP2&1180.31& 150.248&672.93 \\
\hline
BP3&708.56& 59.00&53.86\\
\hline
 \hline
\end{tabular}
 \caption{Cross-sections (in fb) of Higgs productions ($h_{2,3} h_i=1,2,3$)
associated with $Z$ boson at the LHC with $E_{cm}=14$ TeV for the benchmark points. }
\label{crossHZ}
\end{center}
\end{table}

Table~\ref{brh1} and Table~\ref{brHiggses} present the decay branching fractions of $h_1$, $h_2$ and $h_3$, respectively. From  Table ~\ref{brh1} we can see that for all the three benchmark points the lightest Higgs, $h_1$ mostly decays to $\tau\tau$ and $bb$. In case of BP3, we have two additional decay modes $WW$ and $ZZ$. Similar to $h_1$, we can see from Table~\ref{brHiggses} that $h_2$ also mainly decays to  $\tau\tau$ and $bb$. There is also a possibility to decay into $h_1Z$, the branching fractions of which are rather small. In case of BP2 and BP3 $h_2\to h_1h_1$ has small but non-zero branching fraction. Table~\ref{brHiggses} shows that the heaviest neutral Higgs $h_3$ mainly decays to  $h_1$ pair for BP1 and BP2. In case of BP3 it decays to $\tau$ and $b$ pairs like $h_1$.


\renewcommand{\arraystretch}{1.5}
\begin{table}
\hskip -60pt 
\begin{center}
\begin{tabular}{||c||c|c|c|c||}
\hline
\hline
Benchmark&\multicolumn{4}{|c|}{$h_1$ decays}\\
\hline
 points  &$b \bar{b}$&$\tau\bar{\tau}$&$WW$&$ZZ$  \\
 \hline\hline
BP1&0.70 &0.29&-&- \\
\hline
BP2&0.67 &0.32&-&- \\
\hline
BP3&0.67 &0.076&0.14&0.017\\
\hline
 \hline
\end{tabular}
 \end{center}
\caption{The dominant branching fractions of the lightest Higgs boson $h_1$ for the benchmark points.}\label{brh1}
\end{table}

\renewcommand{\arraystretch}{1.5}
\begin{table}
\hskip -60pt 
\begin{center}
\begin{tabular}{||c||c|c|c|c||c|c|c|c||}
 \hline\hline
 \multicolumn{9}{|c|}{Branching fraction}\\
\hline
Benchmark&\multicolumn{4}{|c|}{$h_2$ decays}&\multicolumn{4}{|c|}{$h_3$ decays}\\
\hline
 points  &$b\bar{b}$&$\tau\bar{\tau}$& $h_1 Z$&$h_1 h_1$&$b\bar{b}$&$\tau\bar{\tau}$& $h_1 Z$&$h_1 h_1$   \\
 \hline\hline
BP1&0.68&0.316&1.0$\times 10^{-4}$&- &0.01& 8.7$\times 10^{-3}$&1.6$\times 10^{-5}$&0.98 \\
\hline
BP2&0.62&0.36&1.0$\times 10^{-3}$&0.01& 6.7$\times 10^{-3}$ &8.2$\times 10^{-3}$&3.4$\times 10^{-4}$&0.98 \\
\hline
BP3&0.79&0.19&3.1$\times 10^{-4}$ &1.4$\times 10^{-3}$&0.79&0.19&1.2$\times 10^{-4}$&1.9$\times 10^{-3}$\\
\hline

 \hline
\end{tabular}
 \end{center}
\caption{The dominant branching fractions of heavier Higgs bosons($h_2$ and $h_3$) for the benchmark points.}\label{brHiggses}
\end{table}

\section{Collider phenomenology}

In this study, {\tt CalcHEP} \cite{calchep} is used to calculate the cross-sections, the decay branching fractions and also to generate the events. The couplings and mass spectra are originally generated from the program {\tt CPsuperH2.2} \cite{cpsuph2.0} which is used by CalcHEP via calling the program {\tt CPsuperH2.2}. The standard {\tt CalcHEP-PYTHIA} interface \cite{calpyth}, which uses the SLHA  interface \cite{slha} was then used to pass the {\tt CalcHEP}-generated events to {\tt PYTHIA} \cite{pythia}. Furthermore, all relevant decay information is generated with  {\tt CalcHEP} and is passed to {\tt PYTHIA} through the same interface.  All these  are required since there is no public implementation of CP violating MSSM in {\tt PYTHIA}. Subsequent decays of the produced particles, hadronization and the collider analyses are done with {\tt PYTHIA (version 6.4.5)}. 

We use {\tt CTEQ6L} parton distribution function (PDF) \cite{Lai:1999wy,Pumplin:2002vw}.  In {\tt CalcHEP} we opted for the lowest order $\alpha_s$ evaluation, which is appropriate for a lowest order PDF like {\tt CTEQ6L}. The renormalization/factorization scale in {\tt CalcHEP} is set at $\sqrt{\hat{s}}$. This choice of scale results in a somewhat conservative estimate for the event rates.

For hadronic level simulation we have used {\tt Fastjet-3.0.3} \cite{fastjet} algorithm for the jet formation with the following criteria:
\begin{itemize}
  \item the calorimeter coverage is $\rm |\eta| < 4.5$ 
 
  \item $ p_{T,min}^{jet} = 20$ GeV and jets are ordered in $p_{T}$
  \item leptons ($\rm \ell=e,~\mu$) are selected with
        $p_T \ge 20$ GeV and $\rm |\eta| \le 2.5$
  \item no jet should match with a hard lepton in the event
   \item $\Delta R_{lj}\geq 0.4$ and $\Delta R_{ll}\geq 0.2$ 
  \item Since efficient identification
of the leptons is crucial for our study, we required, on top of the above set of
cuts, that hadronic activity within a cone of $\Delta R = 0.3$ between two isolated leptons should be $\leq 0.5 p^{\ell}_T$ GeV in the specified cone.
\end{itemize}

\begin{figure}[thb]
\begin{center}
\hskip -15pt
\subfigure[\hskip -30pt]{{\epsfig{file=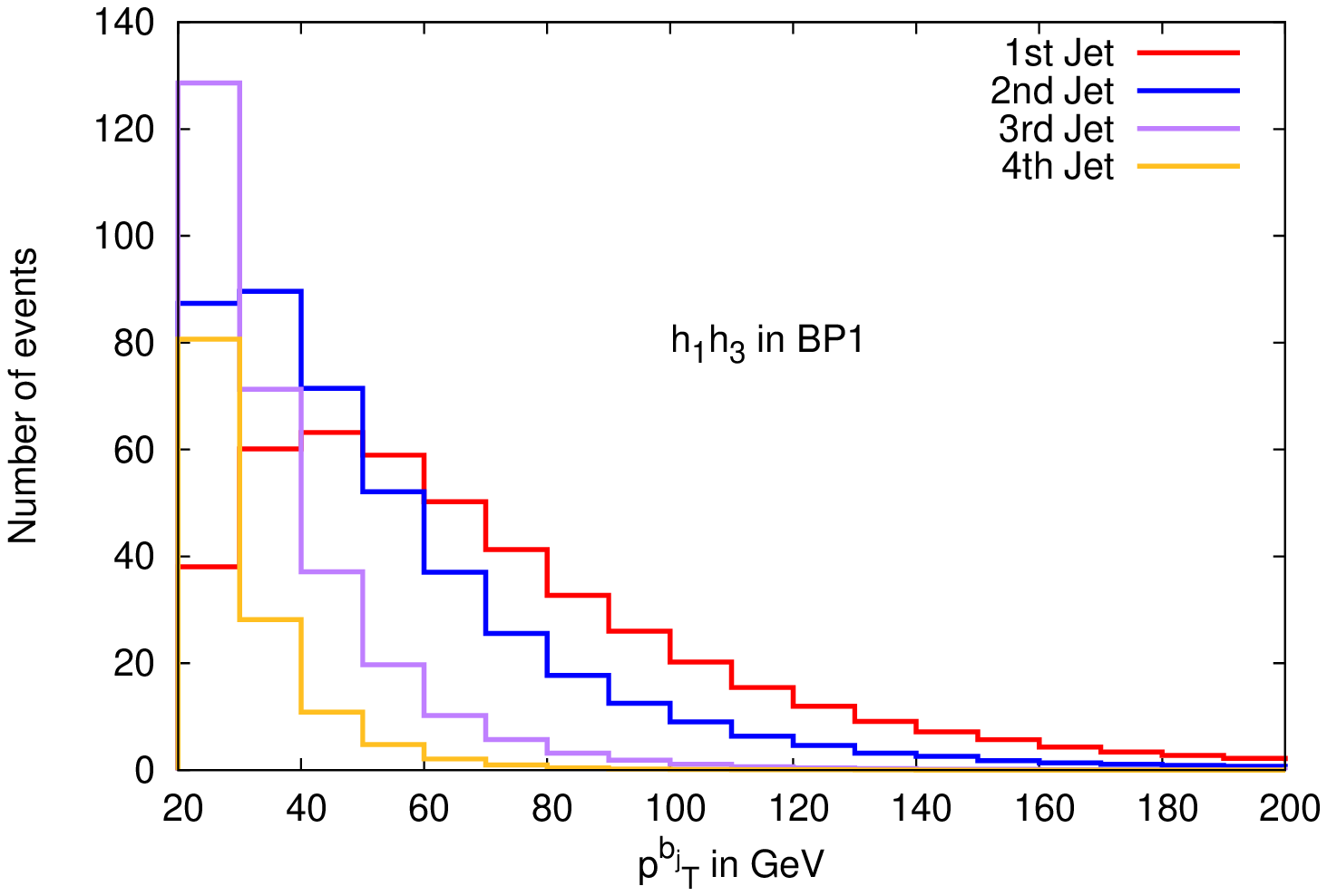,width=6.5 cm,height=7.5cm,angle=-90}}}
\hskip -12pt 
\subfigure[\hskip -30pt]{{\epsfig{file=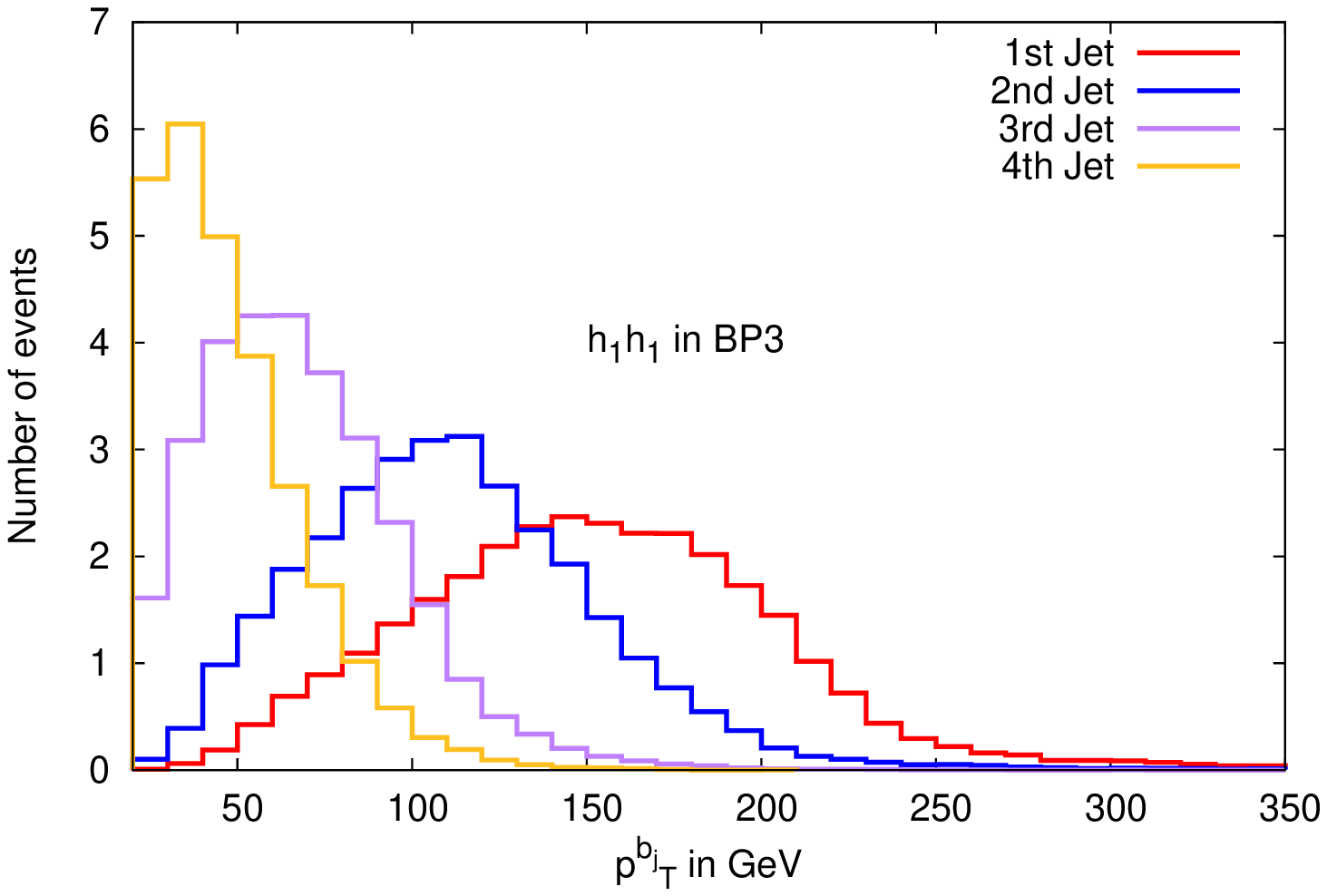,width=6.5cm,height=7.5cm,angle=-90}}}
\caption{ $p^{b_{jet}}_T$ distribution for $h_1h_3$ (a) for BP1 and for  $h_1h_1$ (b) for BP3 at an integrated luminosity of $\cal{L}$ = 10 fb$^{-1}$.}\label{bptdis}
\end{center}
\end{figure}
In the CP-violating scenario, $h_1$ decays dominantly into $b\bar{b}$ and $\tau\bar{\tau}$ (see Table~\ref{brh1}) for all the benchmark points as discussed in the earlier section. In cases of BP1 and BP2 where the light Higgs $h_1$ is relatively light ($<60$ GeV), $b$-quarks lead to soft jets and the $b$-tagging efficiency is small. To illustrate this, we present in Figure~\ref{bptdis} the ordered $p_T$ distributions for $b$-jets coming from $h_1h_3$ for BP1 and from $h_1h_1$ for BP3. We see that for BP1, the lowest $p_T$ $b$-jet can be very soft, $p_T\leq 40$ GeV. For this analysis we have required a $b$-jet tagging efficiency ($\geq$ 50\%) \cite{btag}.
\begin{figure}[thb]
\begin{center}
{\epsfig{file=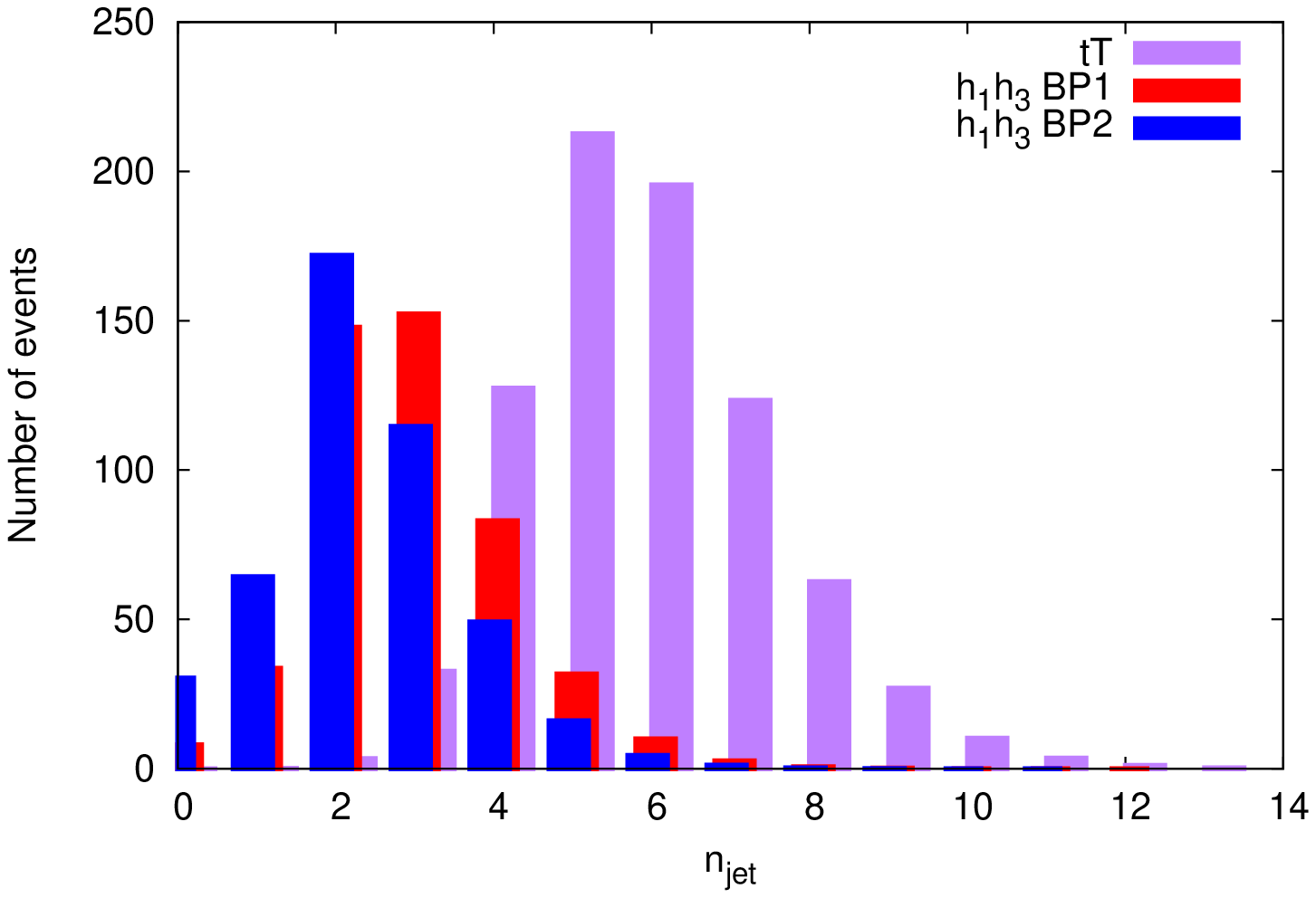,width=6.0 cm,height=8.0cm,angle=-90}}
\caption{Jet multiplicity distributions for $h_1h_3$ for BP1, BP2 and $t\bar{t}$ at an integrated luminosity of $\cal{L}$ = 10 fb$^{-1}$.}
\label{jetm}
\end{center}
\end{figure}

Next we study the jet-multiplicity distribution for $h_1h_3$ for BP1, BP2 and the dominant background $t\bar{t}$. We can see from Figure~\ref{jetm} that the two Higgs production has fewer jets than the $t\bar{t}$. Demanding $n_{jets} \leq 4$ removes most of the $t\bar{t}$ background events. Thus it could be a very useful tool to kill the SM background as well as the SUSY cascade backgrounds which usually have a large number of jets.

\begin{figure}[thb]
\begin{center}
\subfigure[\hskip -30pt]{{\epsfig{file=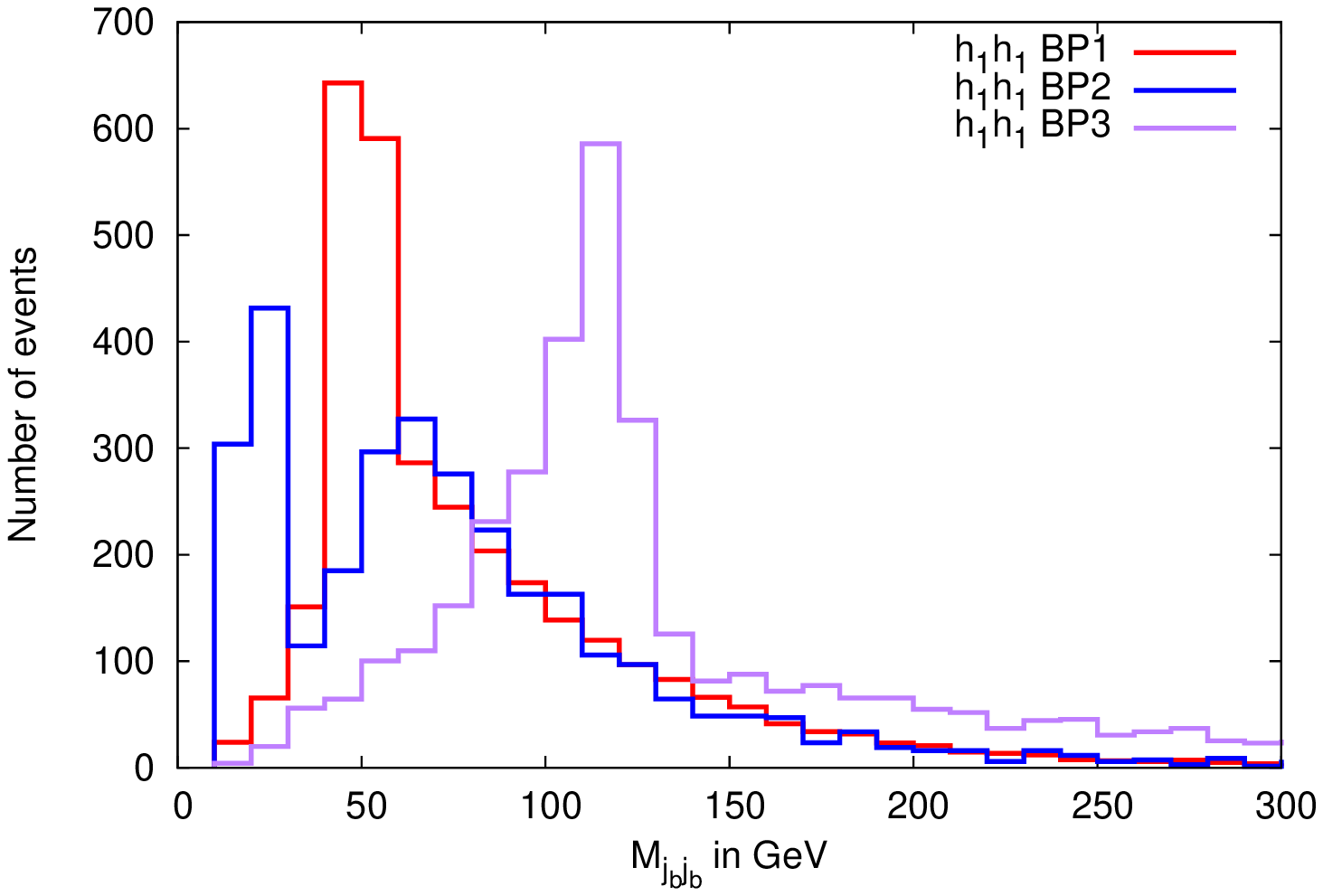,width=7.0 cm,height=7.5cm,angle=-90}}}
\subfigure[\hskip -30pt]{{\epsfig{file=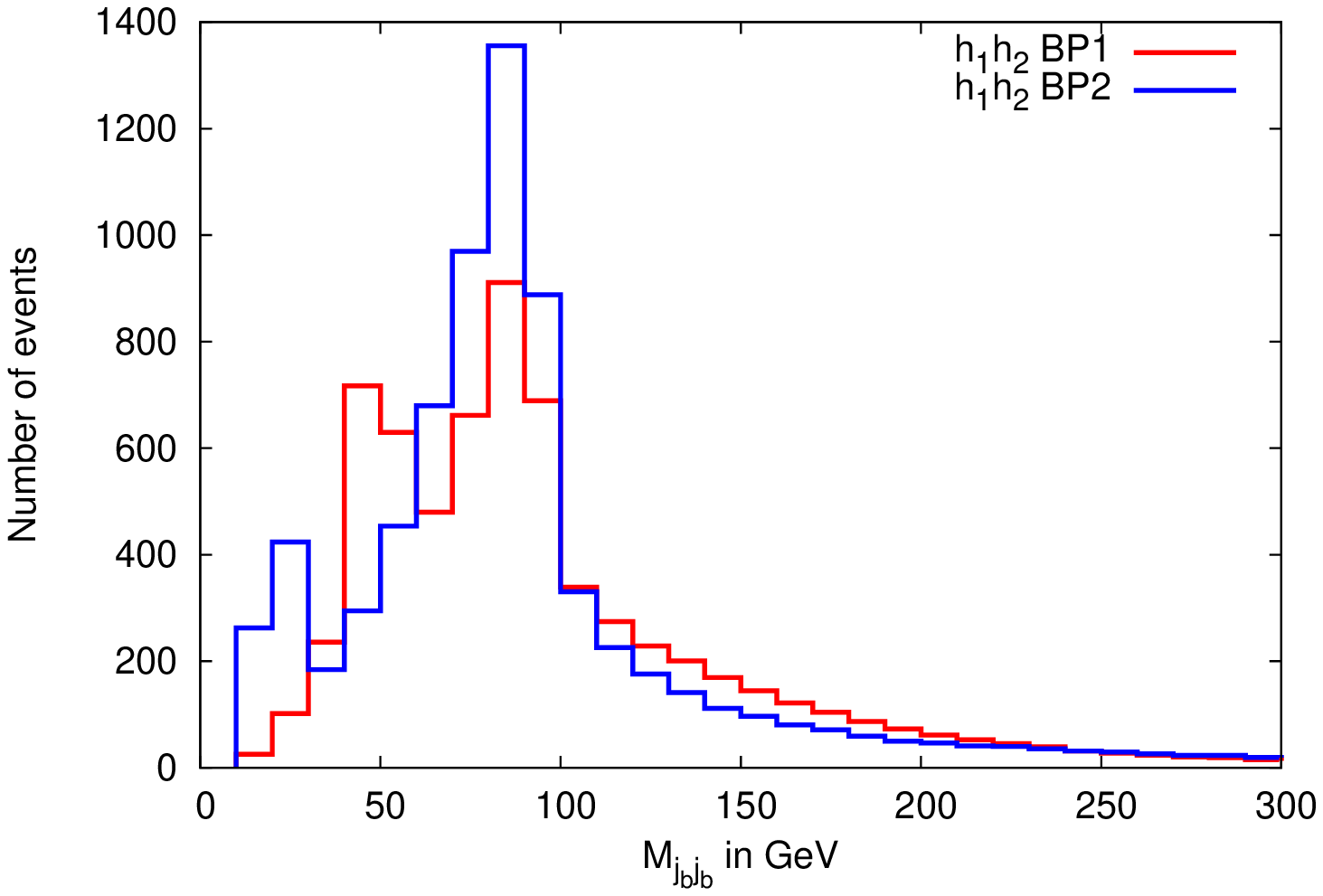,width=7.0 cm,height=7.5cm,angle=-90}}}
\subfigure[\hskip -30pt]{{\epsfig{file=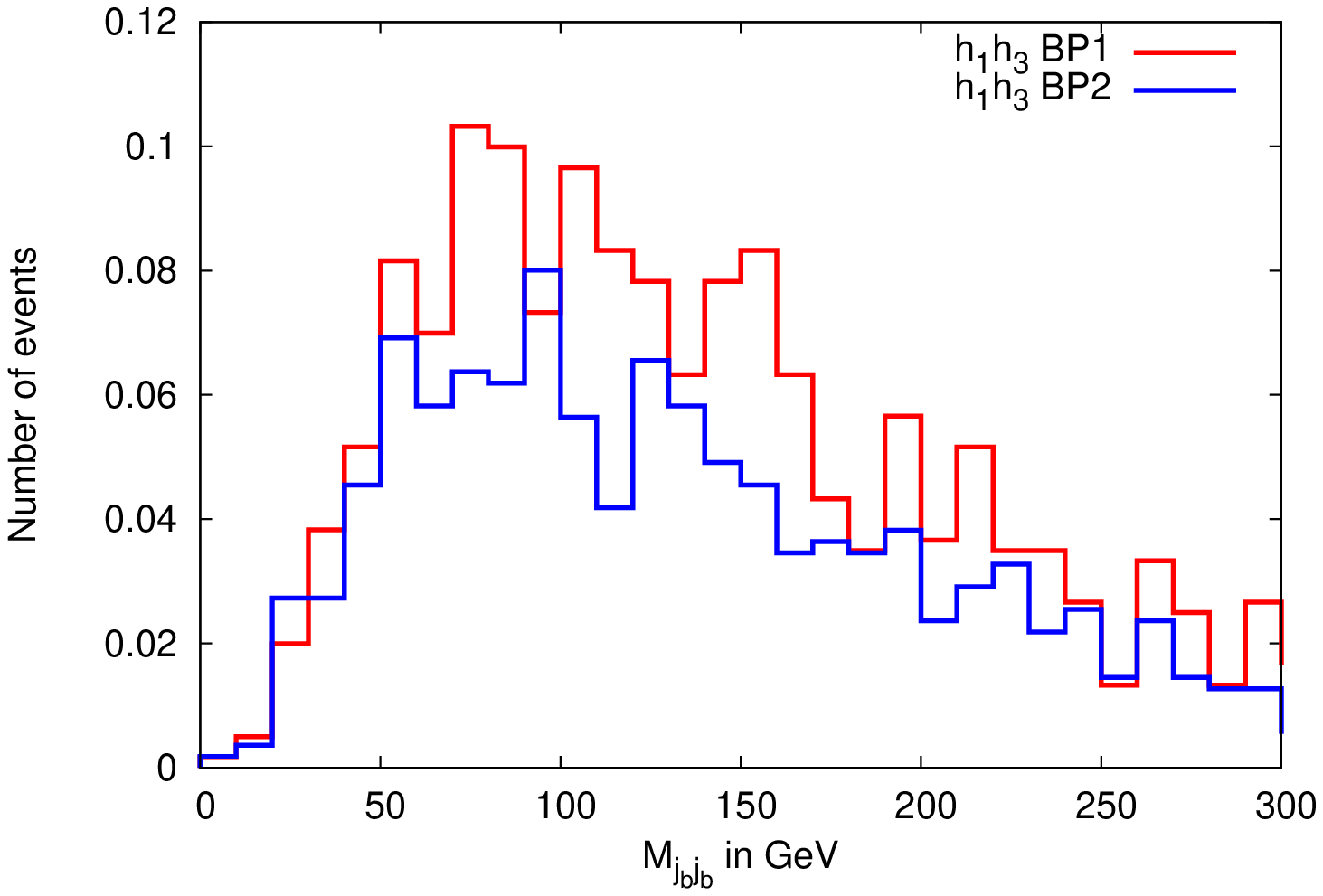,width=7.0 cm,height=7.5cm,angle=-90}}}
\caption{$b$-jet invariant mass distribution coming (a) from $h_1h_1$, (b) from $h_1h_2$ and (c) from $h_1h_3$ for benchmark points at an integrated luminosity of $\cal{L}$ = 10 fb$^{-1}$.}\label{bbinv}
\end{center}
\end{figure}
To see the status of the $b$-final states we first check the $b$-jet invariant mass.
 Figure~\ref{bbinv} shows the invariant mass of two $b$-jets which satisfy the above 
mentioned criteria at an integrated luminosity of 10 fb$^{-1}$ at the LHC with center of mass energy of 14 TeV. In Figure~\ref{bbinv}(a) the $b$-jet invariant mass
 comes from $h_1h_1$ signal for the three benchmark points. The lightest Higgs boson
 peaks are visible for all three benchmark points. In Figure~\ref{bbinv}(b) we show 
both the lightest Higgs $h_1$ peak as well as the second lightest Higgs peak $h_2$, which come from $h_1h_2$ for BP1 and BP2\footnote{For BP3 the number of events are not enough to plot the $b$-jet pair invariant mass distribution}. Similarly Figure~\ref{bbinv}(c) describe the $b$-jet pair invariant mass distribution for BP1 and BP2 coming from $h_1h_3$ signal. Due to small cross-section the mass resolutions are not clear unlike the other two production channels.

\begin{figure}[htb]
\begin{center}
\subfigure[\hskip -30pt]{{\epsfig{file=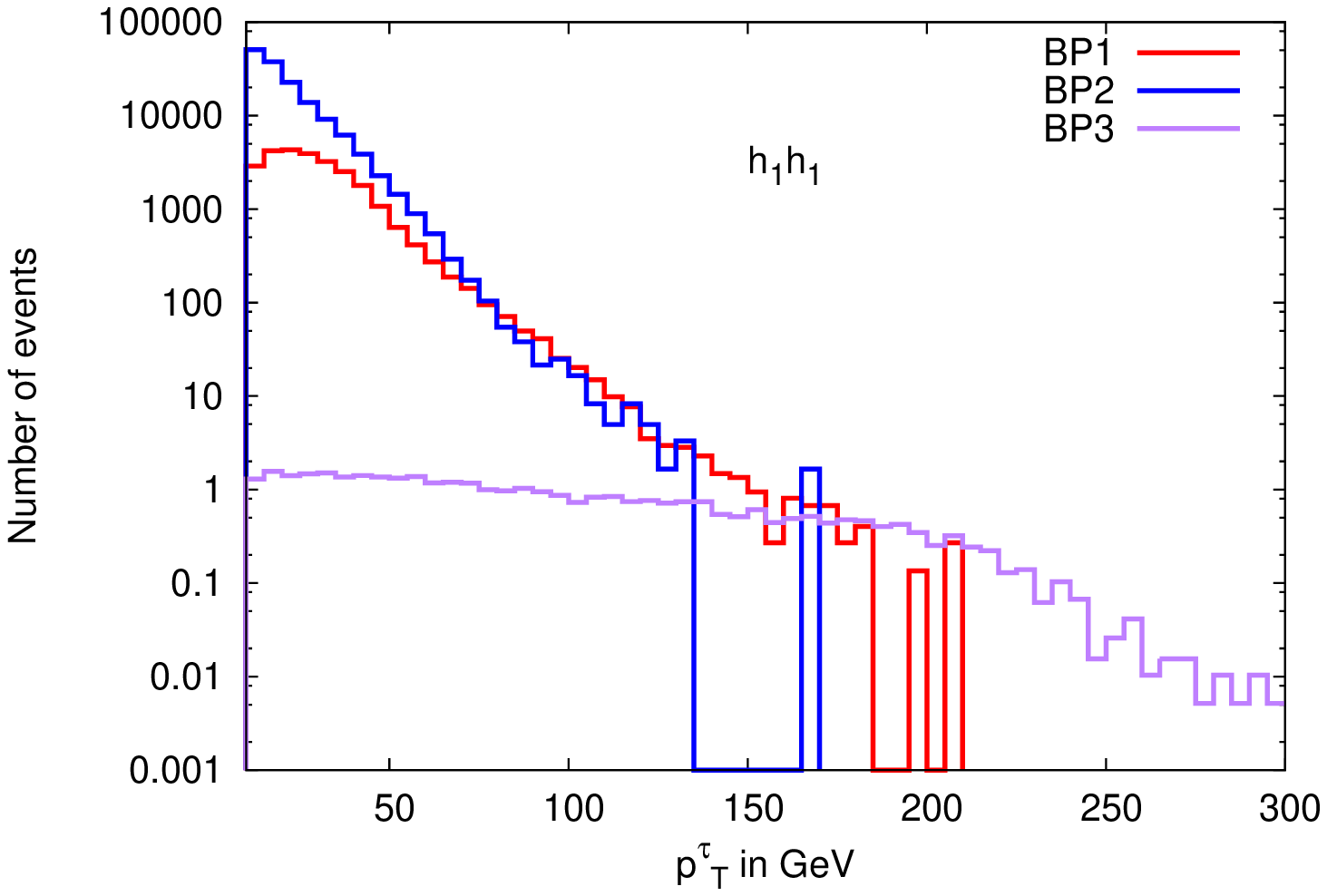,width=7.0 cm,height=7.0cm,angle=-90}}}
\subfigure[\hskip -30pt]{{\epsfig{file=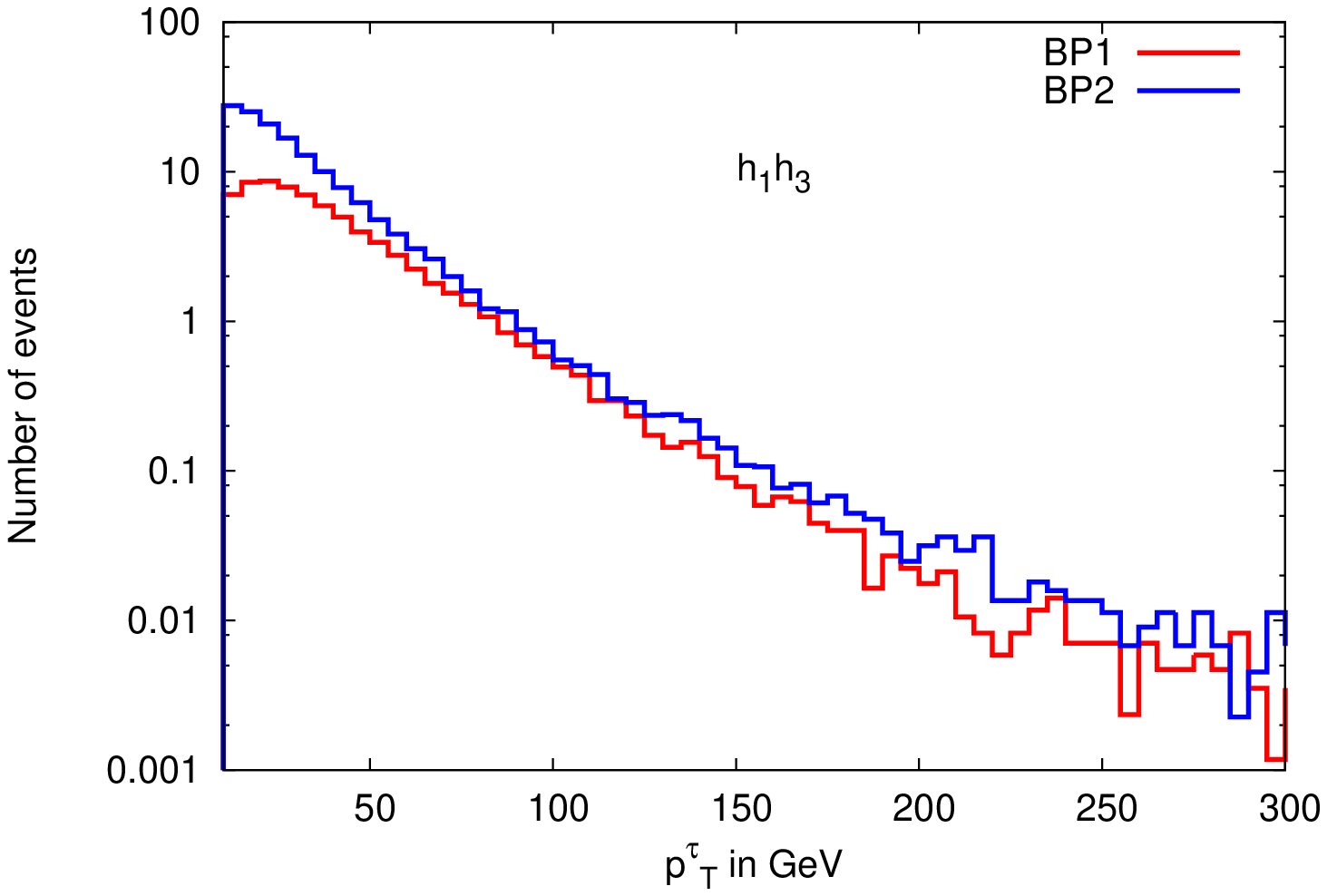,width=7.0 cm,height=7.0cm,angle=-90}}}
\caption{$p_T$ distribution of partonic $\tau$ coming from $h_1h_1$ and $h_1h_3$ for the benchmark points at an integrated luminosity of $\cal{L}$ = 10 fb$^{-1}$. Due to small production cross-section of 
$h_1h_3$ for BP3, the generated events are not enough for the distribution.}\label{pttau}
\end{center}
\end{figure}

All three Higgs bosons decay to $\tau$ pairs with branching fraction  
$\sim 8-30\%$ except for $h_3$  whose branching fraction to $\tau$
pairs is $\mathcal{O}(10^{-3})$ for BP1 and BP2 (see Table~\ref{brh1} and Table~\ref{brHiggses}). For a very light Higgs, in cases of BP1 and BP2, the taus coming from
$h_1$ can be very soft. Boost of the light Higgs ($h_1$)
 of course increases the $p_T$ of taus. Figure \ref{pttau}(a)
shows the $p_T$ distribution of the partonic $\tau$ coming
from $h_1h_1$ production channel. We can clearly see $\tau$s
coming from $h_1$ decay for BP1 and BP2 will have enough boost to
tag them as tau-jet. Figure \ref{pttau}(b) shows that in case of 
$h_1h_3$ production channel the boost of $\tau$s increases further.

Taus coming from Higgs then decay to pions through one prong
or/and three prong decay. In the present study, we use the one-prong (one charged track) hadronic decays of the $\tau$-leptons
 which have a collective branching fraction of about 50\% of which
almost 90\% is comprised of final states with $\pi^\pm$, $\rho$ and
$a_1$ mesons. 
To establish a jet as a $\tau$-jet we take the following approach.
We first check, for each jet coming out of {\tt Fastjet} within $|\eta| \le 2.5$,
if there is a partonic $\tau$ within a cone of $\Delta{R}\le 0.4$ about
the jet-axis. If there is one, then we further ensure that there is
a single charged track within a cone of $\Delta{R} \le 0.1$ of the same
jet axis. This marks a narrow jet character of a $\tau$-jet. Of course
there is an efficiency associated to such kind of a geometric requirement
which is a function of $p_T$ of the concerned jet and has been demonstrated
in the literature \cite{tau1,tau2}. 

\begin{figure}
\begin{center}
\subfigure[\hskip -30pt]{{\epsfig{file=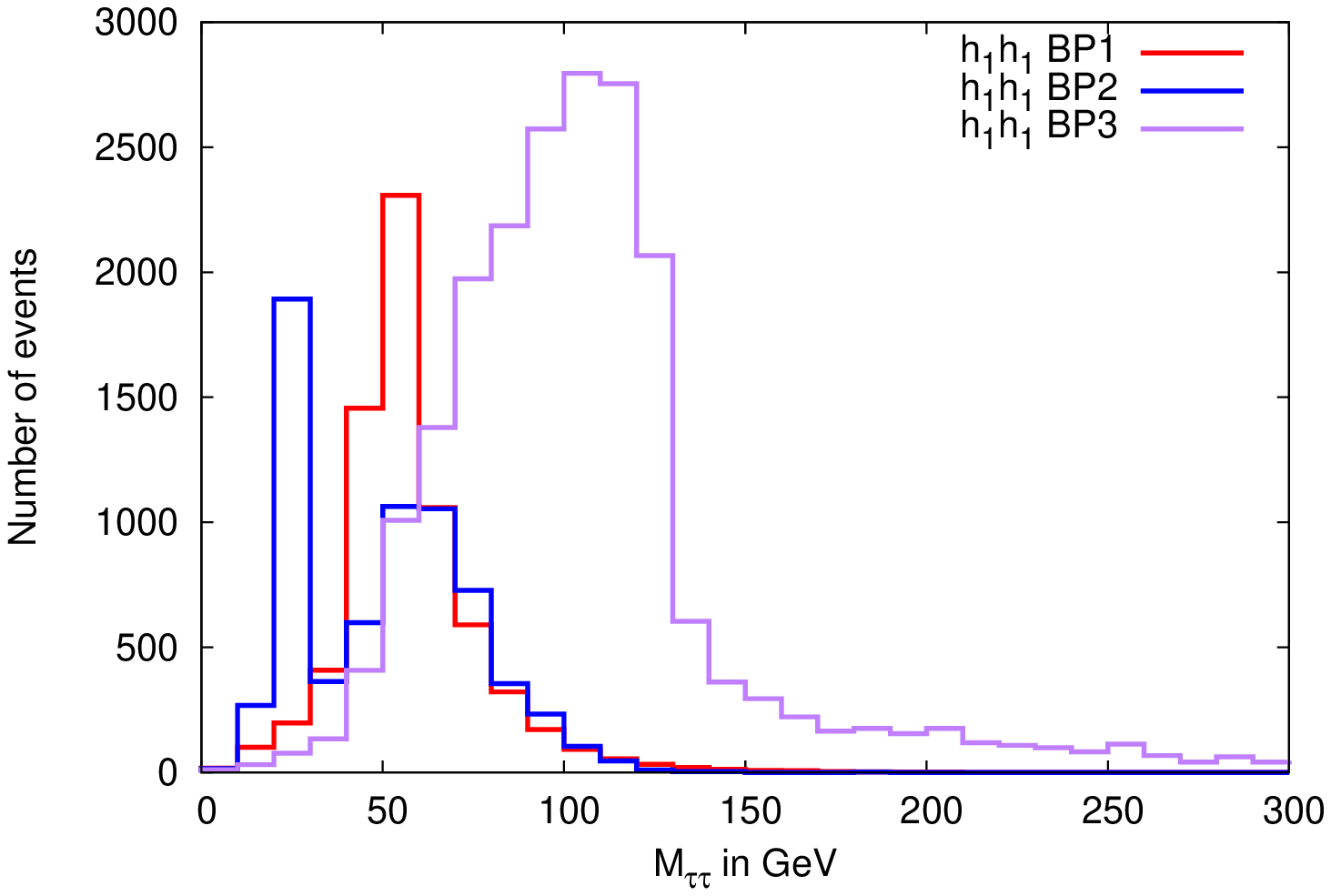,width=7.0 cm,height=7.5cm,angle=-90}}}
\subfigure[\hskip -30pt]{{\epsfig{file=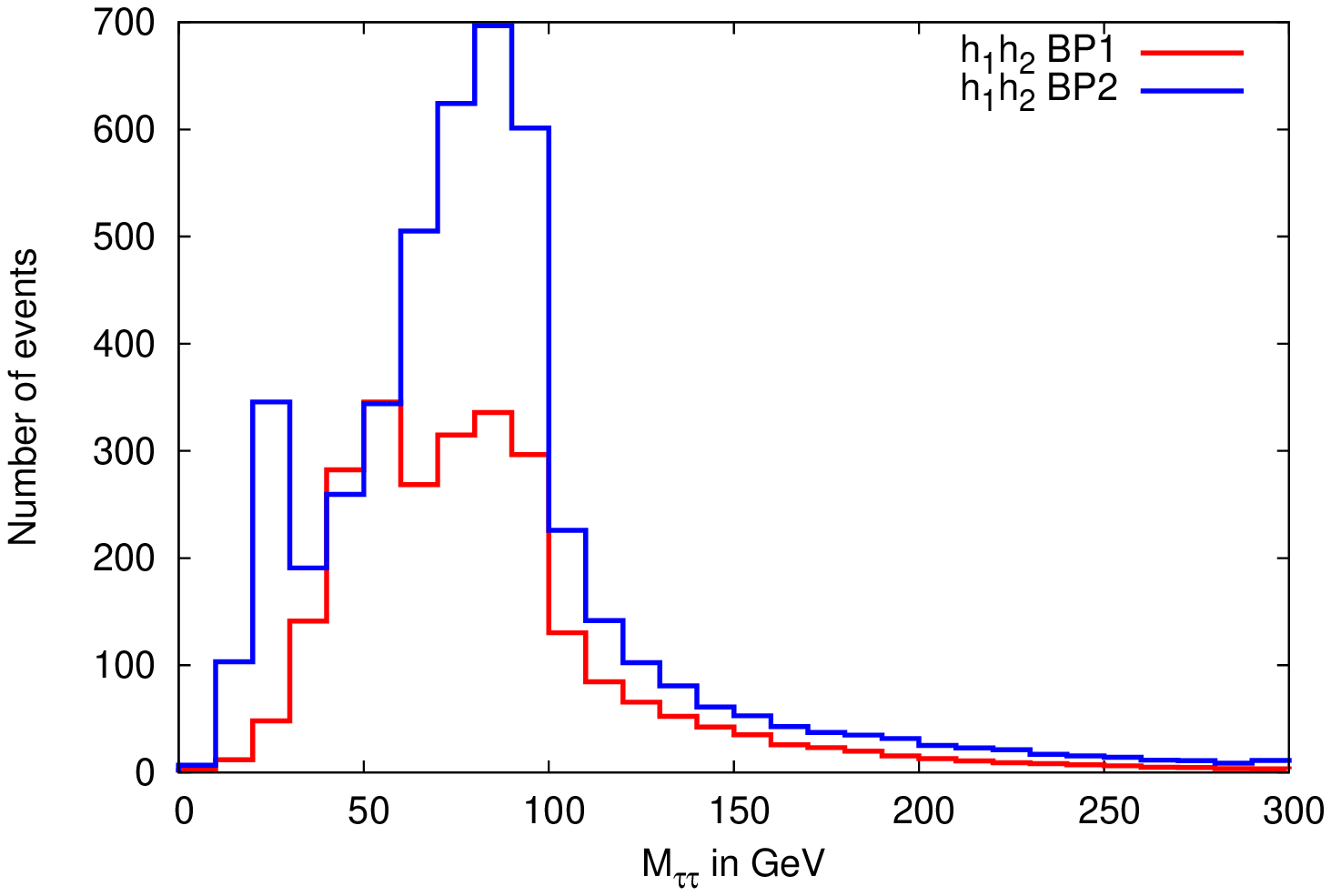,width=7.0 cm,height=7.5cm,angle=-90}}}
\subfigure[\hskip -30pt]{{\epsfig{file=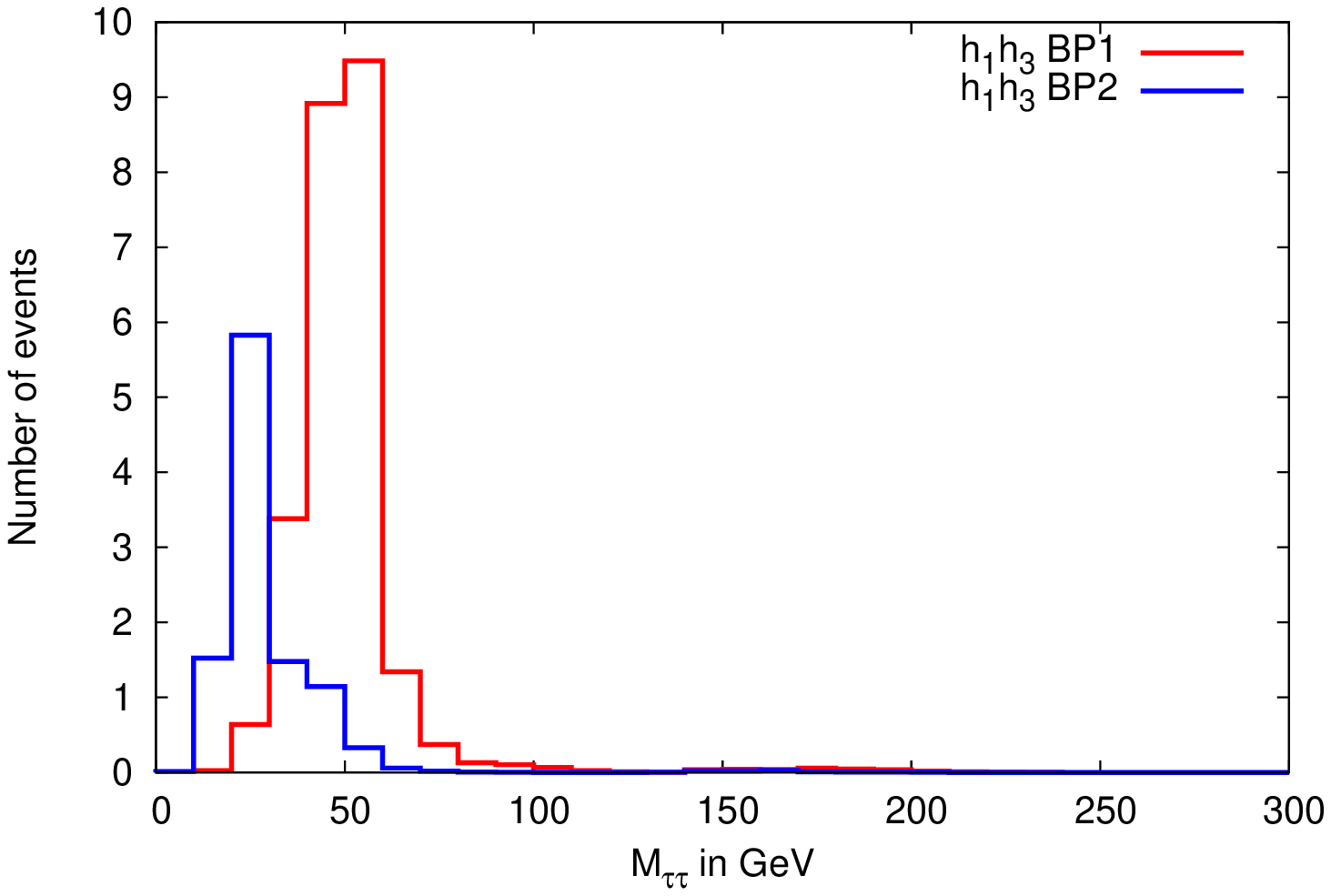,width=7.0 cm,height=7.5cm,angle=-90}}}
\caption{$\tau$-jet invariant mass distribution coming (a) from $h_1h_1$, (b) from $h_1h_2$ and (c) from $h_1h_3$ for benchmark points at an integrated luminosity of $\cal{L}$ = 10 fb$^{-1}$.}\label{ttinv}
\end{center}
\end{figure}

Next we study the the $\tau$ final state by plotting the hadronic $\tau$-jet invariant mass. In Figure~\ref{ttinv} we plot the invariant mass distribution of two hadronic $\tau$-jets coming from the Higgs boson decay. In Figure~\ref{ttinv}(a) the contribution comes from $h_1h_1$ production as before and it is easily seen that the lightest Higgs mass peaks  are much clearer than the $b$-jet invariant mass distributions. Figure~\ref{ttinv}(b)\& (c) show the contribution coming from the production of $h_1h_2$ and $h_1h_3$, respectively for BP1 and BP2. In case of Figure~\ref{ttinv}(c) the $h_3$ mass peak is not visible as $h_3$ mostly decays to $h_1$ pair for both BP1 and BP2 (see table~\ref{brHiggses}).

So far we have seen that the final states with $b$ and $\tau$-jets could be prompted for all the benchmark points if we consider the production channels, i.e., $h_ih_j, i,j=1,2,3$. Let us first discuss the final states with $b$ and $\tau$.

\subsection{Sig1: $3b+2\tau$}
The final state $3b+2\tau$ is possible when at least one Higgs is heavy, i.e.,  $h_2$ or $h_3$, which decays to $Z\, h_1$ or $h_1h_1$. If at least one lightest Higgs $h_1$ decays to tau lepton pair and the other $h_1$ or $Z$ decays to $b$ pairs then we have $4b+2\tau$ final state. This scenarios is possible for BP1 and BP2:

\begin{eqnarray}
pp&\to & h_1h_{2,3}, \nonumber\\
&\to & h_1Zh_1(\rm{or}\,h_1h_1) \to 4b +2\tau .
\label{Sig1a}
\end{eqnarray}

The $b$-tagging efficiency is around 50\%, so tagging 4 $b$-jets will bring 
down the number of signal events. This is the reason we study the $3b +2\tau$
 final state. We choose the final state as:

$$\rm{sig1:}\, n_{jets}\leq 5+( \geq 3 b-jet) +\,( \geq \,2\tau-jet \,) + ( \ptmiss \leq\, 30\, \rm{GeV}).$$

We consider $t\bar{t}$, $t\bar{t}Z$, $t\bar{t}W$, $ZZ$ and $t\bar{t}b\bar{b}$ as the main SM backgrounds.  Table \ref{3b2t} presents the number of events for 
sig1 for signal and backgrounds at an integrated luminosity of 10 fb$^{-1}$. Table \ref{3b2t} also presents the numbers of events with window cut of $\pm 10$ GeV around the mass peak of respective invariant mass distribution. Here we give the number of events with the window cuts around $h_1$, $h_2$ and $h_3$ for both $bb$ and $\tau\tau$ invariant mass distribution. Determination of these mass peaks depends on the relative number of signal events over background events.

\renewcommand{\arraystretch}{1.0}
\begin{table}[bht]
\begin{center}
\begin{tabular}{||c||c|c|c||c|c|c|c|c||}
\hline
\hline
Signal &\multicolumn{3}{|c|}{ Benchmark Points }&\multicolumn{5}{|c|}{ Backgrounds }\\
 &BP1&BP2&BP3&$t\bar{t}$&$t\bar{t}Z$&$t\bar{t}W$&$ZZ$&$t\bar{t}b\bar{b}$\\
\hline
sig1&52.30&21.60&0.80&1.00&0.00&0.00&0.80&0.06\\
\hline
sig1+$|m_{bb}-m_{h_1}|\leq 10$ GeV&25.30&8.80&0.20&0.00&0.00&0.00&0.00&0.00\\
 \hline
sig1+$|m_{bb}-m_{h_2}|\leq 10$ GeV&3.70&2.10&0.00&0.10&0.00&0.00&0.00&0.00\\
 \hline
sig1+$|m_{bb}-m_{h_3}|\leq 10$ GeV&1.80&0.12&0.00&0.00&0.00&0.00&0.00&0.00\\
 \hline
\hline
sig1+$|m_{\tau\tau}-m_{h1}|\leq 10$ GeV&33.70&19.00&0.09&0.00&0.00&0.00&0.00&0.00\\
 \hline
sig1+$|m_{\tau\tau}-m_{h2}|\leq 10$ GeV&33.70&19.00&0.09&0.00&0.00&0.00&0.60&0.00\\
 \hline
sig1+$|m_{\tau\tau}-m_{h3}|\leq 10$ GeV&0.07&0.20&0.00&0.00&0.00&0.00&0.00&0.00\\
 \hline
\hline
\end{tabular}
 \caption{Number of events after the selection cuts for sig1 final states for the benchmark points and backgrounds at an integrated luminosity of 10 fb$^{-1}$ at the LHC with $E_{cm}=14$ TeV. }
\label{3b2t}
\end{center}
\end{table}
From Table~\ref{3b2t} we see that Sig1 has 7.1$\sigma$ significance over background for BP1 at an integrated luminosity of 10 fb$^{-1}$. For BP2 and BP3 it is 4.5$\sigma$ and $0.5\sigma$, respectively. For $h_1$ peak we can get $\geq 5\sigma$ significance for BP1 for both $bb$ and $\tau\tau$ invariant mass distribution. The corresponding numbers for BP2 are $3\sigma$ and $4.4\sigma$ respectively. In case of $h_2$ and $h_3$ peak, for a comparable signal significance over background one needs to go for higher luminosity.

Next we consider the case when two of the $h_1$ decay to tau pairs 
and the final state is $(\geq 2 b-jet\,)+\,( \geq \,4\tau-jet)$.  The decay
branching fraction of $h_1\to \tau \tau$ is around 30\% which is much lower
than the $h_1\to b\bar{b}$ branching fraction. The tau coming from such a
light Higgs ($h_1$) is of low $p_T$ which reduces the $\tau$ detection efficiency. Because of these two effects the final state does not have many events at 10 fb$^{-1}$ integrated luminosity.

There is a possibility that the heavier Higgses ($h_{2,3}$) decay to 
$h_1Z$ in the case of $h_1h_{2,3}$ productions, which leads to $h_1h_1Z$.
Similarly $h_{2,3}Z$ production also leads to the above final state when $h_{2,3}\to h_1h_1$ . When $Z$ decays to lepton pair and if we tag only $3b$ then it can give final state like $3b+OSD+(|m_{\ell\ell}-M_Z|\leq 5\, \rm{GeV})+(\ptmiss\leq 30\, \rm{GeV})$, where $OSD$ corresponds to opposite sign dilepton.

Comparing the production cross-sections from Table~\ref{crossH} and decay branching fractions from Table~\ref{brh1} and Table~\ref{brHiggses}, we see that the contribution from $h_1h_{2,3}$ production would be negligible due to low $h_{2,3}\to h_1 Z$ branching fraction for the chosen benchmark points. On the other hand $h_{3}Z$ has relatively large production cross section at least for BP1 and BP2 but fails to contribute due to demand of $3b$ tagging coming from very light Higgs ($h_1$). One needs to go for very high luminosity to look for this final state.

\subsection{Sig2: $2b+2\tau$}
Unlike for the other benchmark points, in BP3, $h_3\to h_1 h_1 $ is very small,
and $h_3$ mostly decays to $b$ or tau pairs. Thus the final state with $2b+2\tau$ looks promising. Thus both the heavy($h_{2,3}$) and light ($h_1$) Higgs bosons can decay either to $b$ pair or $\tau$ pair which leads to $2b+2\tau$ final state. If the $b$s and $\tau$s are coming from the heavier Higgs ($h_{2,3}$), then 
they have a high $p_T$. On the other hand when they come from the light Higgs
($h_1$) they have a very low $p_T$. We study the final state as:

$$\rm{sig2:}n_{jets}\leq 5+ (\geq 2 b-jet) +\, (\geq \,2\tau-jet \,) + (\ptmiss \leq\, 30\, \rm{GeV}).$$

Table~\ref{2b2t} presents the number of events for the signal and backgrounds 
at an integrated luminosity of 10 fb$^{-1}$. We can see that sig2 has 13.5 $\sigma$, $10\sigma$ and $0.6\sigma$ significance with 10 fb$^{-1}$ of luminosity for BP1, BP2 and BP3, respectively. This could be a useful channel to look for the light Higgs scenarios. We then put a window cut in the $b\bar{b}$ invariant mass distribution around the light Higgs mass peak ($m_{h_1}$) as $|m_{bb}-m_{h_1}|\leq 10$ GeV.  The signal significance for this case does not change much from the previous one and it is $12\sigma$ and $10.4\sigma$ for BP1 and BP2. The buried Higgs scenarios can be probed at the LHC.  Even when we put the window cut around the next mass peak, i.e., $|m_{bb}-m_{h_2}|\leq 10$ GeV, the signal significance for BP1 and BP2 still remains around $5\sigma$ at an integrated luminosity of 10 fb$^{-1}$. For heavier Higgs mass peak resolution, i.e for $m_{h_3}$ one needs to go to higher luminosity, at least to 43 fb$^{-1}$ of luminosity in the case of BP1 and BP2. We also investigate the scenario where we take window cuts around $\tau\tau$ invariant mass peak. In this case the reach for the Higgs mass peaks is possible in relatively higher luminosity. 
\renewcommand{\arraystretch}{1.0}
\begin{table}[bht]
\begin{center}
\begin{tabular}{||c||c|c|c||c|c|c|c|c||}
\hline
\hline
Signal &\multicolumn{3}{|c|}{ Benchmark Points }&\multicolumn{5}{|c|}{ Backgrounds }\\
 &BP1&BP2&BP3&$t\bar{t}$&$t\bar{t}Z$&$t\bar{t}W$&$ZZ$&$t\bar{t}b\bar{b}$\\
\hline
sig2&501.30&350.80&19.00&812.10&0.30&0.50&57.70&0.20\\
\hline
&&&&65.00&0.04&0.05&6.20&0.00\\
sig2+$|m_{bb}-m_{h_1}|\leq 10$ GeV&195.00&129.00&4.00&23.70&0.00&0.00&0.60&0.00\\
&&&&59.00&0.05&0.05&0.60&0.00\\
 \hline
&&&&103.00&0.01&0.08&15.00&0.06\\
sig2+$|m_{bb}-m_{h_2}|\leq 10$ GeV&69.00&56.00&0.00&104.10&0.01&0.08&16.00&0.06\\
&&&&1.0&0.00&0.00&0.00&0.00\\
 \hline
&&&&60.00&0.04&0.06&0.30&0.00\\
sig2+$|m_{bb}-m_{h_3}|\leq 10$ GeV&22.00&8.20&0.00&60.00&0.04&0.06&0.30&0.00\\
&&&&1.00&0.00&0.00&0.00&0.00\\
 \hline
\hline
sig2+$|m_{\tau\tau}-m_{h1}|\leq 10$ GeV&0.10&0.00&0.00&0.00&0.00&0.00&0.00&0.00\\
 \hline
&&&&101.00&0.04&0.10&17.00&0.06\\
sig2+$|m_{\tau\tau}-m_{h2}|\leq 10$ GeV&52.00&33.00&0.20&103.00&0.04&0.10&17.00&0.06\\
&&&&1.00&0.00&0.00&0.00&0.00\\
 \hline
&&&&105.00&0.01&0.07&0.30&0.06\\
sig2+$|m_{\tau\tau}-m_{h3}|\leq 10$ GeV&4.00&3.00&0.10&104.00&0.03&0.07&0.20&0.00\\
&&&&1.00&0.00&0.00&0.00&0.00\\
 \hline
\hline
\end{tabular}
 \caption{Number of events after the selection cuts for sig2 final states for the benchmark points and backgrounds at an integrated luminosity of 10 fb$^{-1}$ at the LHC with $E_{cm}=14$ TeV. The different rows of background events for a given column correspond to BP1, BP2 and BP3, respectively as they differ depending on the window cuts around the mass peaks. }
\label{2b2t}
\end{center}
\end{table}

\subsection{Sig3: $2\ell$}
In this section we will see the exclusive leptonic final states, i.e., the 
final states with $\mu$ and $e$. Though the branching fractions of Higgses to
lepton pair are very small, these tiny branching fractions can be crucial for precision measurement of  invariant mass peak. The leptonic channel is particularly handy when it comes to determination of a very small Higgs mass ($\lesssim 50$ GeV).
\begin{figure}
\begin{center}
\subfigure[]{{\epsfig{file=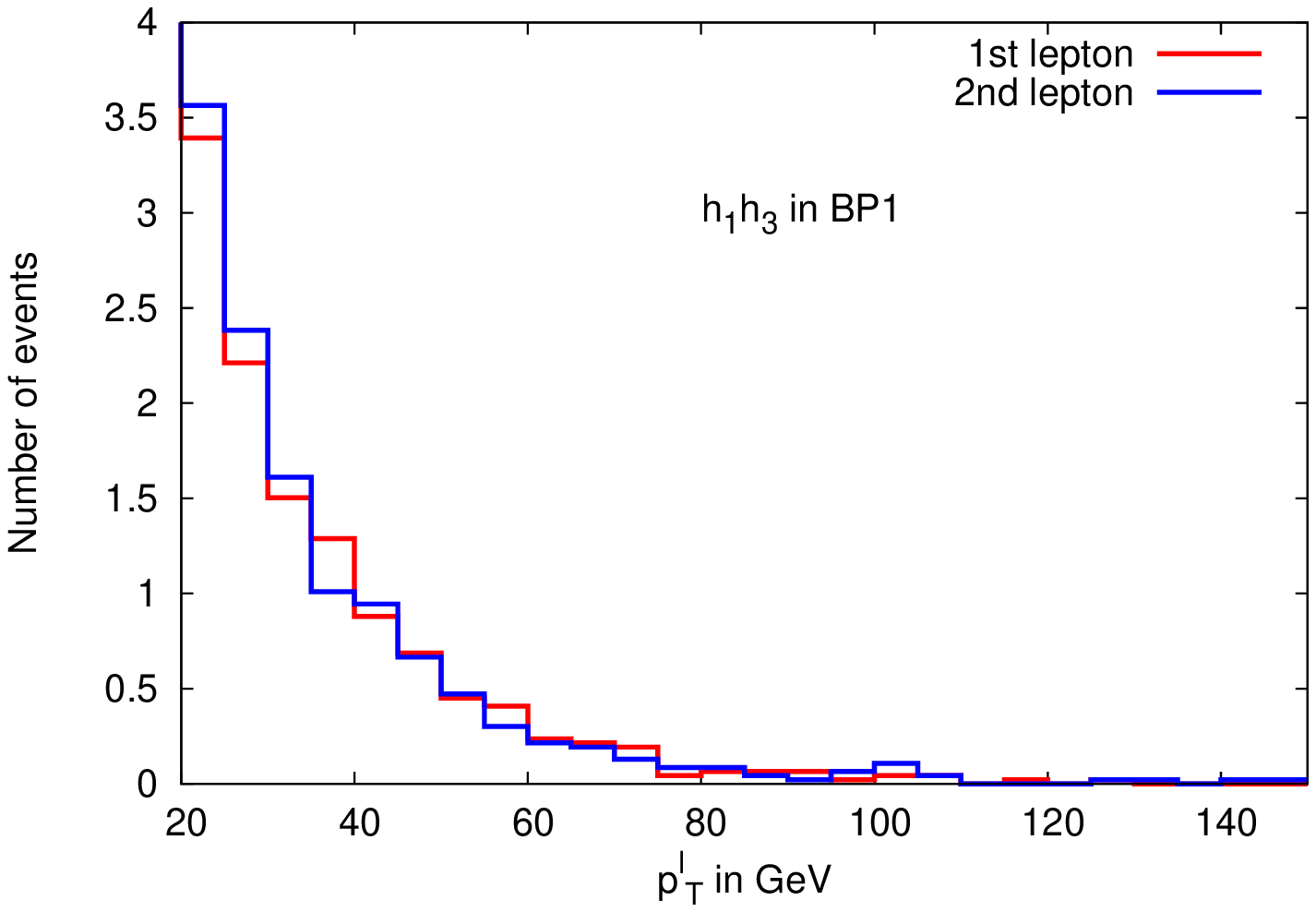,width=6.5 cm,height=7.5cm,angle=-90}}}
\hskip -12pt 
\subfigure[]{{\epsfig{file=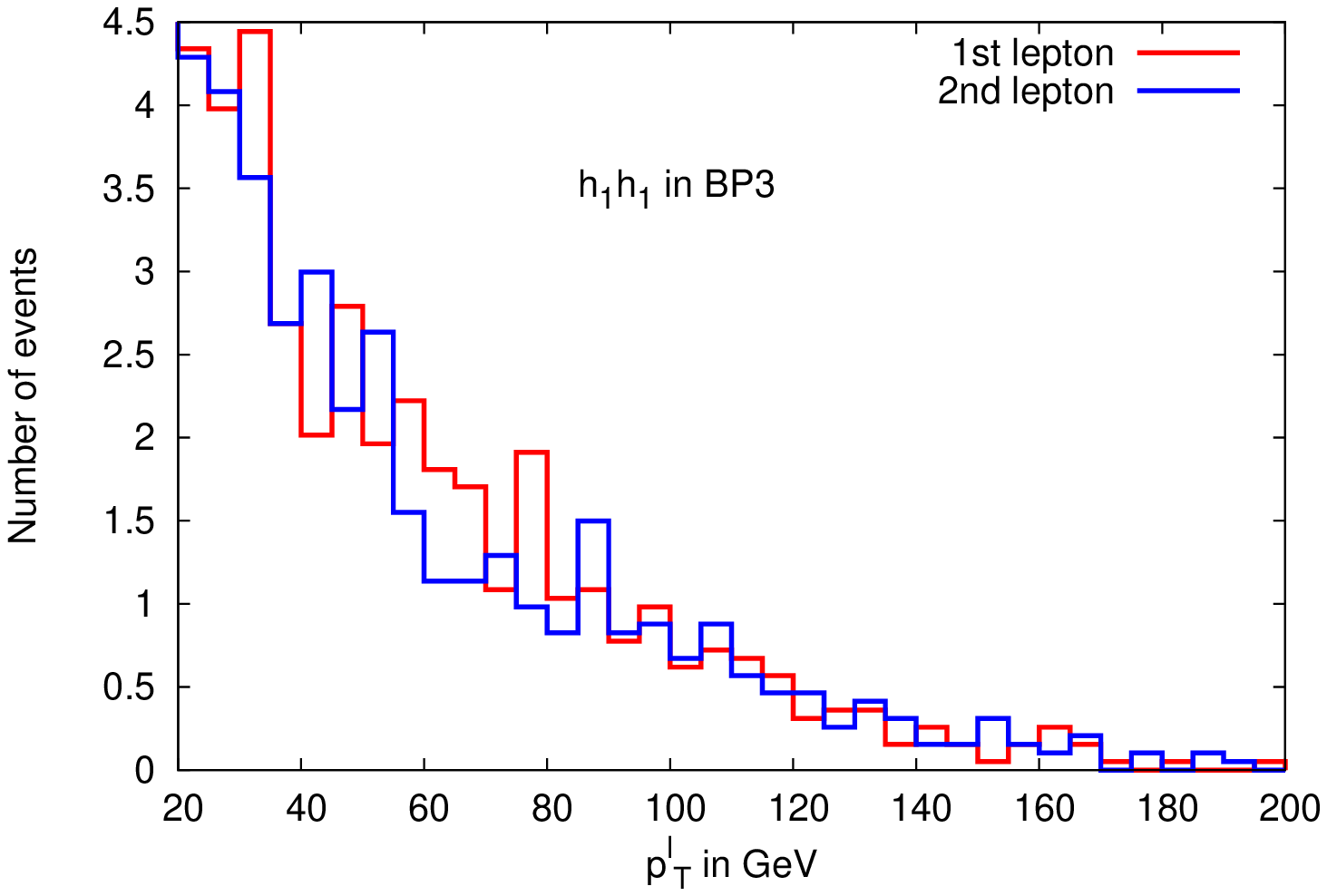,width=6.5 cm,height=7.5cm,angle=-90}}}
\caption{$p^{\ell}_T$ distribution from (a) $h_1h_3$ for BP1 and from (b)$h_1h_1$ for BP3.}\label{leppt}
\end{center}
\end{figure}
In Figure~\ref{leppt} the lepton $p_T$ distribution comes from (a) $h_1h_3$ for BP1 and (b) $h_1h_1$ for BP3. Clearly leptons for BP1 can be treated as hard leptons ($p_T\geq 20$ GeV) but for BP3 they can be 
as hard as 200 GeV.

\begin{figure}
\begin{center}
\subfigure[\hskip -30pt]{{\epsfig{file=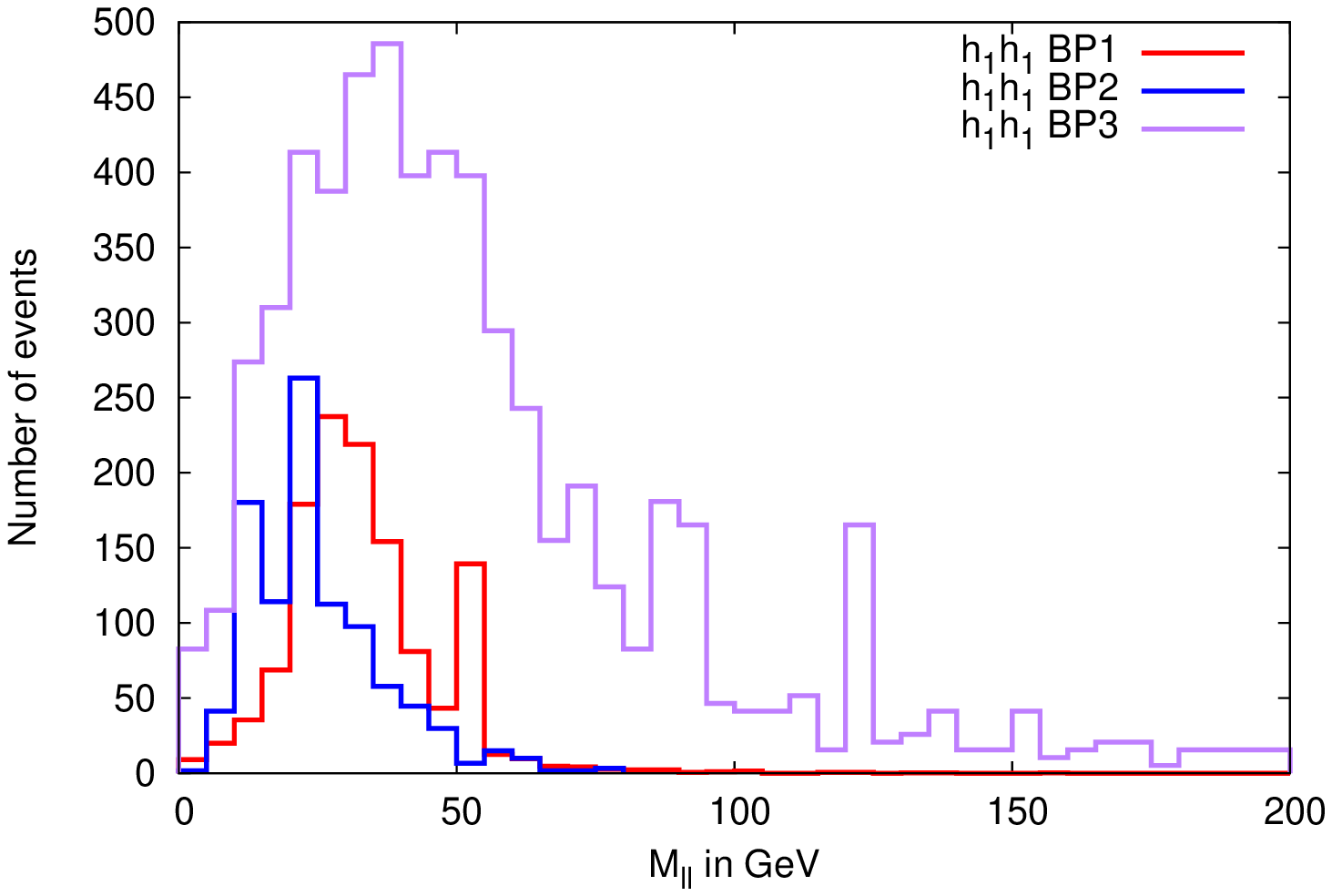,width=7.0 cm,height=7.5cm,angle=-90}}}
\subfigure[\hskip -30pt]{{\epsfig{file=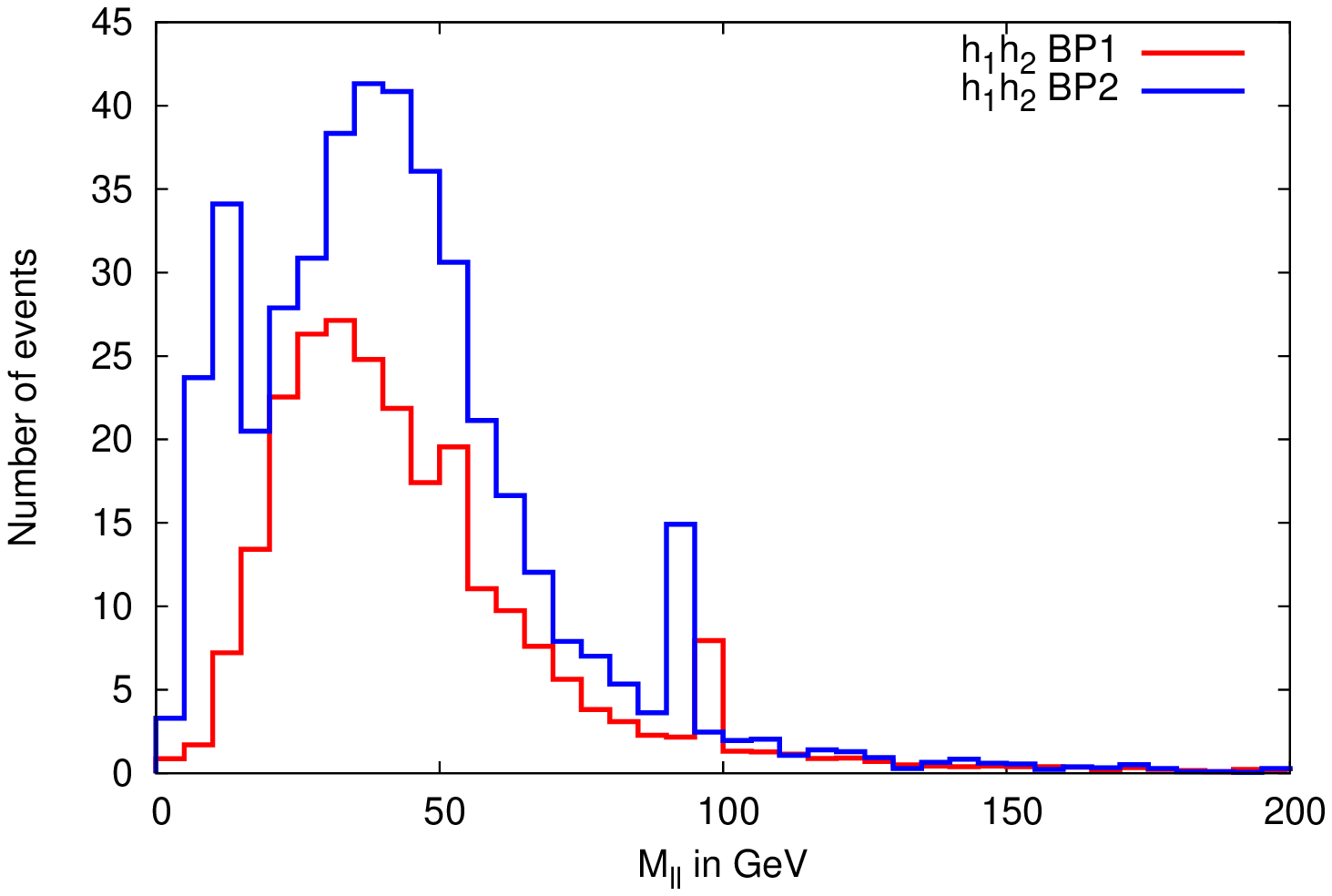,width=7.0 cm,height=7.5cm,angle=-90}}}
\caption{Lepton invariant mass distribution coming (a) from $h_1h_1$, (b) from $h_1h_2$  for benchmark pints at an integrated luminosity of $\cal{L}$ = 10 fb$^{-1}$.}\label{llinv}
\end{center}
\end{figure}

Figure~\ref{llinv} describes the dilepton invariant mass coming (a) from $h_1h_1$ and (b) from $h_1h_2$. From Figure~\ref{llinv}(a) we can see the $h_1$ peaks for all three benchmark points. On the other hand Figure~\ref{llinv}(b) shows 
both the Higgs mass peaks, i.e., $h_1$  around 30 and 50 GeV, $h_2$ around 95 GeV. 

We first analyse the dilepton final states which could be interesting 
in determining the very light Higgs scenario with precision. We define the final state as: $\rm{sig3:} \,\,2\ell$. Table~\ref{2l} presents the number of events
for the final state sig3 for both signal and backgrounds at an integrated luminosity of 10 fb$^{-1}$. The dominant background events are coming from $t\bar{t}$ and gauge boson pair production ($VV$). The signal significance 
for the dilepton final state (sig3) reaches $5\sigma$ for BP2 only. The signal
significance for BP1 crosses $3\sigma$ at 10 fb$^{-1}$ of luminosity. For the light Higgs ($h_1$) mass peak the significance is $7.6\sigma$ for BP2 at 10 fb$^{-1}$ of luminosity but for other benchmark points one needs higher luminosity. Specially to determine the Higgs mass peak for BP3 a very high luminosity is needed. 

Next we also investigated the $4\ell$ final state where both the Higgses decay
into lepton pairs. The prospect for this final state does not look promising 
at least for low luminosity and with 14 TeV LHC.

\renewcommand{\arraystretch}{1.0}
\begin{table}
\begin{center}
\begin{tabular}{||c||c|c|c||c|c|c|c|c||}
\hline
\hline
Signal &\multicolumn{3}{|c|}{ Benchmark Points }&\multicolumn{5}{|c|}{ Backgrounds }\\
 &BP1&BP2&BP3&$t\bar{t}$&$t\bar{t}Z$&$t\bar{t}W$&$VV$&$t\bar{t}b\bar{b}$\\
\hline
sig3: $2\ell$ &200.00&370.00&15.00&2007.00&52.00&43.00&1590.50&3.00\\
\hline
&&&&485.00&6.00&7.80&28.00&0.40\\
sig3+$|m_{\ell\ell}-m_{h_1}|\leq 10$ GeV&20.00&205.00&0.60&491.00&4.00&5.00&15.00&0.70\\
&&&&53.00&3.60&4.20&5.00&0.10\\
 \hline
&&&&153.00&5.00&6.40&706.00&0.20\\
sig3+$|m_{\ell\ell}-m_{h_2}|\leq 10$ GeV&5.00&5.20&0.00&156.00&5.00&6.40&710.00&0.20\\
&&&&0.00&0.01&0.05&1.20&0.00\\
 \hline
&&&&53.00&3.60&4.30&63.00&0.20\\
sig3+$|m_{\ell\ell}-m_{h_3}|\leq 10$ GeV&0.70&0.50&0.00&53.00&3.60&4.30&62.00&0.20\\
&&&&0.00&0.01&0.03&1.10&0.00\\
 \hline
\hline
\end{tabular}
 \caption{Number of events after the selection cuts for $2\ell$ (sig3) final states for the benchmark points and backgrounds at an integrated luminosity of 10 fb$^{-1}$ at the LHC with $E_{cm}=14$ TeV. The different rows of background events for a given column correspond to BP1, BP2 and BP3, respectively as they differ depending on the window cuts around the mass peaks.}
\label{2l}
\end{center}
\end{table}

So far we have presented the dominant Standard Model (SM) backgrounds that 
contribute to the final states. There are other reducible model backgrounds
which we also have calculated. They are $H^\pm W^\mp$, $H^\pm h_{i=1,2,3}$,
$H^\pm H^\mp$, respectively. We find that their contributions are negligible for
the final states we have considered here. The susy backgrounds and supersymmetric backgrounds associated with charged Higgs production have been addressed
in great detail in \cite{Bandyopadhyay:2010tv} and it is shown that most 
of the time  the final states in supersymmetric cascade decays come
with large number of jets, which is unlike the case here.

\section{Summary and discussion}
From our analysis it is clear that the Higgs pair production
is interesting in spite of being electroweak production
process. We have studied various possible final states that could 
come from the two Higgs productions. For some signal topologies an integrated
 luminosity of 10 fb$^{-1}$ is enough to reach $5\sigma$ of
 signal significance. Specially $2b+2\tau$ (Sig2) final state looks promising.  
 We have seen that it is also possible to reconstruct the Higgs mass peak,
both via $bb$ invariant mass and through $\tau\tau$ invariant mass distribution.

We have also studied the leptonic final state which also has a great prospect
due to its precision measurement possibility and can come handy for light Higgs 
mass discovery. The signal topologies coming from Higgs pair 
productions are very different from CP-conserving case due to the 
existence of the light buried Higgs. LHC at 14 TeV has a good chance
to explore this possibility once it starts taking data. With more data
coming in one can look for $bb\tau\tau$ or $bb\ell\ell$ invariant mass
which can determine the heavy Higgs ($h_{2,3}$) mass peak and also one can distinguish $h_iZ$ events from $h_ih_j$ events.

\vspace*{1cm}

\noindent
{\bf Acknowledgments:}

The authors acknowledge support from the Academy of Finland (Project No 137960).
PB wants to thank Helsinki Institute of Physics for the visit
during the early stages of the project and KIAS overseas travel grant. PB also thanks Prof. Jae Sik Lee for useful discussions.

\end{document}